\documentclass[reprint,superscriptaddress,showpacs,showkeys,amsmath,amssymb,floatfix,aps,pra,longbibliography]{revtex4-1}

\usepackage{graphicx}
\usepackage{bm}

\usepackage[dvipsnames]{xcolor}
\usepackage{physics}
\usepackage{placeins}
\usepackage{graphics}
\usepackage{color}
\usepackage{hyperref}
\usepackage{multirow}
\usepackage{blindtext}
\usepackage{gensymb}
\usepackage{textcomp}
\begin{document}


\title{Tuning proximity spin-orbit coupling in graphene/NbSe$_2$ heterostructures via twist angle}

\author{Thomas Naimer}
\email{thomas.naimer@physik.uni-regensburg.de}
\affiliation{Institute for Theoretical Physics, University of Regensburg, 93040 Regensburg, Germany}
\author{Martin Gmitra}
\affiliation{Institute of Physics, Pavol Jozef \v{S}af\'{a}rik University in Ko\v{s}ice, 04001 Ko\v{s}ice, Slovakia}
\affiliation{Institute of Experimental Physics, Slovak Academy of Sciences, 04001 Ko\v{s}ice, Slovakia}
\author{Jaroslav Fabian}
\affiliation{Institute for Theoretical Physics, University of Regensburg, 93040 Regensburg, Germany}

\begin{abstract}
We investigate the effect of the twist angle on the proximity spin-orbit coupling (SOC) in graphene/NbSe$_2$ heterostructures from first principles. The low-energy Dirac bands of several different commensurate twisted supercells are fitted to a model Hamiltonian, allowing us to study the twist-angle dependency of the SOC in detail. We predict that the magnitude of the Rashba SOC can triple, when going from $\Theta=0^\circ$ to $\Theta=30^\circ$ twist angle. Furthermore, at a twist angle of $\Theta\approx23^\circ$ the in-plane spin texture acquires a large radial component, corresponding to a Rashba angle of up to $\Phi=25^\circ$. 
The twist-angle dependence of the extracted proximity SOC is explained by
analyzing the orbital decomposition of the Dirac states to reveal with which NbSe$_2$ bands they hybridize strongest.  Finally, we employ a Kubo formula to evaluate the efficiency of  conventional and unconventional charge-to-spin conversion in the studied heterostructures.
\end{abstract}

\maketitle

\section{Introduction}
Combining different 2D materials into one heterostructure held together by van der Waals forces has become a well trodden path of engineering new physics~\cite{Geim2013:Nat,Duong2017:ACS}; in this approach, the materials can imprint some of their own properties onto another by proximity effects. Graphene has been the most researched material in this context, since it provides high electron mobility~\cite{Bolotin2008:SuspendedGrapheneHighmobility1,Du2008:SuspendedGrapheneHighmobility2} and long spin-relaxation times~\cite{Singh2016:APL,Drogeler2016:NL}. Embedding it into a heterostructure can enhance its spin-orbit coupling~\cite{Gmitra2015:TMDCgraphene1,Gmitra2016:TMDCgraphene2,Wang2015:NC,Avsar2014:NC,Naimer2021:paper1,Naimer2023:paper2,Zollner2023:PRB} (e.g. graphene/WSe$_2$), equip it with magnetic properties~\cite{Karpiak2020:GrCGT,Wang2015:PRL,Wei2016:NM,Zhang2015:PRB,Zollner2018:NJP,Zhang2018:PRB} (e.g. graphene/Cr$_2$Ge$_2$Te$_6$) or turn it into a superconductor~\cite{Gani2019:PRB,Moriya2020:PRB,Zhang2020:PRB} (e.g. graphene/NbSe$_2$). 

Although graphene/NbSe$_2$ is mostly studied in the context of superconducting graphene ~\cite{Wang2017:Nat,Xi2016:NP,Khestanova2018:NL,Gani2019:PRB,Moriya2020:PRB,Zhang2020:PRB}, the spin-orbit coupling of graphene is expected  to be enhanced in such heterostructures as well. This is because NbSe$_2$  has a large spin-orbit coupling, similar to semiconducting transition-metal dichalcogenides (TMDCs).
In heterostructures of graphene and semiconducting TMDCs (e.g., WSe$_2$), the influence of the twist angle on the proximity SOC was found to be of great impact: Not only does the magnitude and type of SOC vary, but also a radial component is introduced to the in-plane spin structure for non-zero twist angles, as studied  by tight binding models~\cite{Li2019:TwistTB1,David2019:TwistTB2, Peterfalvi2022:radialrashba} and \emph{ab initio} simulations~\cite{Naimer2021:paper1,Lee22:PRB:radialrashba3,Zollner2023:PRB}. Recently, the emergence of a radial component of the Rashba spin-orbit field was confirmed experimentally ~\cite{yang2024:arxiv:GRTMDCradialexp}.
However, unlike graphene heterostructures based on semiconducting TMDCs which feature Dirac points within the semiconductor band gap, in graphene/NbSe$_2$ the Dirac point lies within the metallic bands of NbSe$_2$. This brings certain challenges for deciphering the proximity effects, making their first-principles investigations useful for understanding the interplay between SOC and superconductivity~\cite{Gani2019:PRB,Moriya2020:PRB,Zhang2020:PRB} and for charge-to-spin conversion in graphene/NbSe$_2$ heterostructures~\cite{Ingla_Aynes2022:expradialrashba2}. Particularly the collinear charge-to-spin conversion (unconventional Rashba Edelstein effect, UREE~\cite{Ingla_Aynes2022:expradialrashba2,Peterfalvi2022:radialrashba,Veneri22:PRB:radialrashba2,Lee22:PRB:radialrashba3,Camosi_2022:2DM:expradialrashba3,Ontonso2023:PRA:expradialrashba4,yang2024:arxiv:GRTMDCradialexp}) as opposed to the more usual perpendicular charge-to-spin conversion (Rashba Edelstein effect, REE~\cite{Offidani2017:PRL,Drydal2014:PRB,Edelstein1990:SSC}) has a high potential for being realized in these heterostructures. Since all SOC parameters (and especially the Rashba phase angle $\Phi$, linked to collinear charge-to-spin conversion) can vary strongly with the twist angle 
as shown for graphene/TMDCs~\cite{Naimer2021:paper1,Lee22:PRB:radialrashba3,Zollner2023:PRB,Li2019:TwistTB1,David2019:TwistTB2,Peterfalvi2022:radialrashba,yang2024:arxiv:GRTMDCradialexp} and graphene/topological insulators~\cite{Zhang14:PRB:TIonlyTB,Song2018:TIGRheteroDFT,Naimer2023:paper2,Kiemle22:acs}, it is important to uncover such dependencies for graphene/NbSe$_2$ heterostructures as well. 

To predict twist-angle dependence of proximity SOC, we perform density functional theory (DFT) calculations on several commensurate graphene/NbSe$_2$ supercells. We fit the obtained band structures around the Dirac points to a model Hamiltonian which comprises different SOC terms. 
The main result of the paper are the extracted SOC values depending on the twist angle   $\Theta$.  
More specifically, we find that valley Zeeman and Rashba SOC both have a magnitude of just under 1~meV for twist angles $0^\circ\leq\Theta\leq15^\circ$. Then, for $15^\circ\leq\Theta\leq30^\circ$, the Rashba SOC steadily increases reaching a maximum of about 2.5~meV at $\Theta=30^\circ$, while the valley Zeeman SOC steadily decreases to zero magnitude at $\Theta=30^\circ$. At $\Theta=22^\circ$, the Rashba phase angle has a maximum with a value of $\Phi=-25^\circ$.
The strong SOC at $\Theta=30^\circ$ twist angle, which is purely of the Rashba type, can have multiple experimental implications. For one, the combination of superconductivity and Rashba SOC, which can be simultaneously achieved by proximity effects in such graphene/NbSe$_2$ heterostructures, can lead to superconducting diode effect\cite{Hu2007:PRL:superconddiode,Baumgartner2022:JP:superconddiode,Baumgartner2022:NN:superconddiode,Jiang2022:NP:superconddiode,Costa2023:PRB:superconddiode}. For another, strong Rashba SOC without the presence of valley-Zeeman SOC offers an ideal platform for charge-to-spin conversion via the (U)REE\cite{Veneri22:PRB:radialrashba2,Drydal2014:PRB,Lee22:PRB:radialrashba3}.

To explore the latter, we explicitly investigate the potential of the different graphene/NbSe$_2$ supercells for charge-to-spin conversion. By using the Kubo formula in the Smrcka-Streda formulation, we evaluate at what twist angles the potential for measuring REE or UREE is the highest.
Our findings support the intuitive picture following from the SOC parameters: the highest yield for REE is possible for $\Theta=30^\circ$ (maximal Rashba SOC, minimal valley-Zeeman SOC). The highest yield for UREE is possible for $\Theta\approx22^\circ$ (maximal Rashba phase angle $\Phi$, high Rashba SOC).

The paper is structured as follows: In Sec.~\ref{Sec:strucs} we show the supercells used for the DFT calculations and the model Hamiltonian used for the fitting. Sec.~\ref{Sec:bandoffsets} discusses the band alignments and the challenges linked to the metallic nature of the NbSe$_2$. In Sec.~\ref{Sec:SOC}, the twist-angle dependence of the proximity SOC parameters is presented. The potential for charge-to-spin conversion is explored in Sec.~\ref{Sec:UREE} within linear response theory. In App.~\ref{App:EVSSOC} and App.~\ref{App:relax} the effects of an external electric field and relaxation are discussed respectively. Finally, in App.~\ref{App:UREEsupp} we present some details on the calculations done in Sec.~\ref{Sec:UREE}.

\section{Structures and methods}
\label{Sec:strucs}
\begin{figure}
    \includegraphics[width=.99\linewidth]{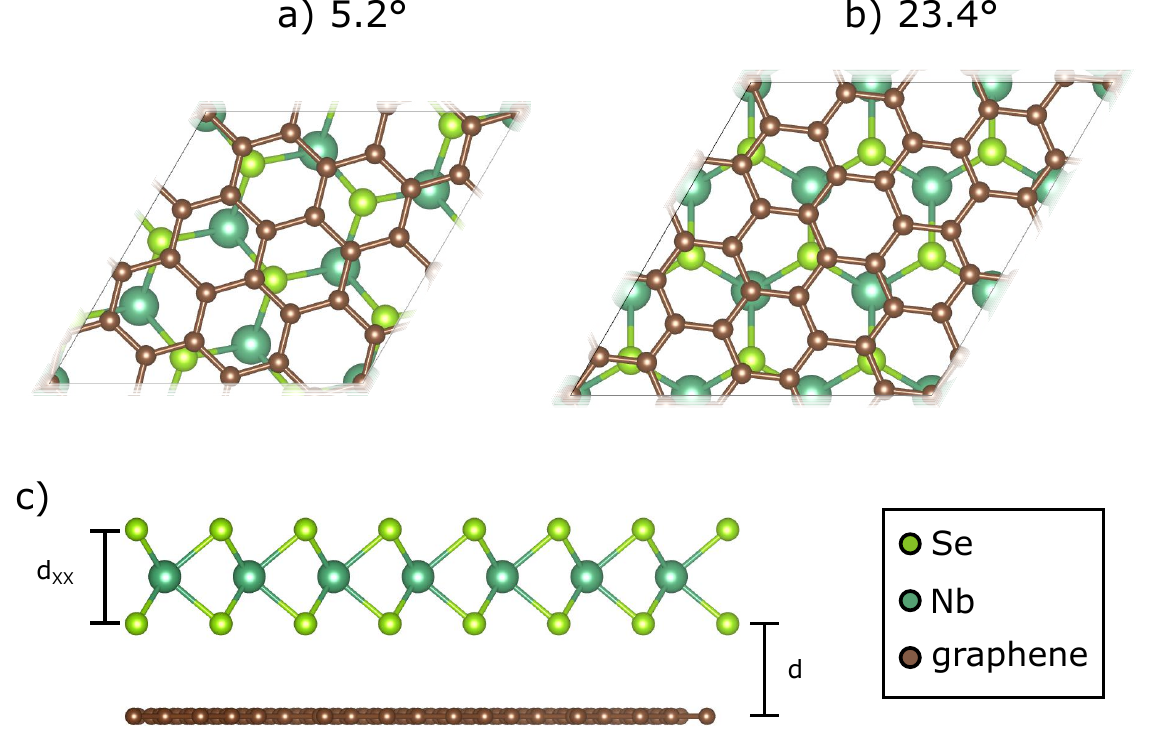}
    \caption{Crystal structure models of graphene/NbSe$_2$ commensurate heterostructures. a)-b) Bottom view of the 5.2$^\circ$ and the 23.4$^\circ$ supercell. c) side view of a heterostructure with interlayer distance $d$ and NbSe$_2$ thickness $d_{XX}$.\label{Fig:structures}}
\end{figure}

Combining monolayer graphene (lattice constant $a_{Gr}=2.46$~\AA) and a monolayer of NbSe$_2$ (lattice constant $a_{NbSe_2}=3.48~$\AA~\cite{Ding2011:PB} and thickness $d_{XX}=3.358~$\AA~\cite{Gani2019:PRB}), we construct the supercells listed in Tab.~\ref{Tab:structures} implementing the coincidence lattice method~\citep{Koda2016:JPCC,Carr2020:NRM}. The chosen interlayer distance (we study the effects of structural relaxation in  App.~\ref{App:relax}) is $d=3.3~$\AA. The integer attributes $(n,m)$ determine the lattice vectors of the (graphene or NbSe$_2$) supercell
\begin{align}
\mathbf{a}^S_{(n,m)}&=n\cdot \mathbf{a}+m\cdot \mathbf{b} \\
\mathbf{b}^S_{(n,m)}&=-m\cdot \mathbf{a}+(n+m)\cdot \mathbf{b},
\end{align}
where $\mathbf{a}$ and $\mathbf{b}$ are the primitive lattice vectors (of graphene or NbSe$_2$). These new supercell lattice vectors in turn determine the twist angle $\Theta$ and strain $\epsilon$ of the heterostructure (for details on this notation, see~\cite{Naimer2021:paper1}). Two examples of such heterostructures are shown in Fig.~\ref{Fig:structures}. Assuming that the lateral degree of shifting plays a minor role for larger supercells (see Ref.~\cite{Naimer2021:paper1,Naimer2023:paper2}), we use the convention of a shifting position where at the corner of the supercell a Nb atom sits on top of a C atom. In order to achieve a commensurate heterostructure suitable for DFT calculations, we need to introduce the strain $\epsilon$ in either the NbSe$_2$ or the graphene. Since graphene's electronic structure is less affected by strain\cite{Si2016:graphenestrain1,Choi2010:graphenestrain2,Gui2008:PRB:graphenestrain3,Grassano2020:PRB:graphenestrainworkfunction}, we choose to put all the strain on graphene and leave the NbSe$_2$ unstrained. Finally, we add a vacuum of 20~\AA~to avoid interactions between periodic images in our slab geometry.
Electronic structure calculations are then performed by density functional theory (DFT)~\citep{Hohenberg1964:PRB} with {\tt Quantum ESPRESSO}~\citep{Giannozzi2009:JPCM}.
Self-consistent calculations are carried out with a $k$-point sampling of $n_k\times n_k\times 1$. The number $n_k$ is listed in Table~\ref{Tab:structures} for all twist angles. We use charge density energy cutoff 350~Ry and wave function energy cutoff 60~Ry for the scalar relativistic pseudopotential with the projector augmented wave method~\citep{Kresse1999:PRB} with the 
Perdew-Burke-Ernzerhof exchange correlation functional~\citep{Perdew1996:PRL}. Graphene's $d$-orbitals are not included in the calculations. We used D-2 van der Waals corrections~\citep{Grimme2006:JCC,Grimme2010:JCP,Barone2009:JCC}.

\begin{table}[]
\caption{Structural information of all investigated NbSe$_2$/graphene heterostructures. We list the integer attributes $(n,m)$ (graphene) and $(n',m')$ (NbSe$_2$), the strain $\epsilon$ imposed on graphene, the twist angle $\Theta$ between graphene and NbSe$_2$ and the number of atoms in the supercell $N_\text{at}$. The $(n,m)$ marked by a star indicates the supercells where $n-m=3\cdot k, k\in \mathbb{Z}$ and therefore the Dirac cone is folded back to $\Gamma$.
The $k$-mesh density $n_k\times n_k\times 1$ used in the self-consistent calculations is also listed.
\label{Tab:structures}}
\begin{tabular}{c|cc|c|c|c}

$\Theta $ &$(n,m)$ & $(n',m')$& $\epsilon $&$N_\text{at}$ &$n_k$ \\
$[\degree]$&&&[\%]&\\
\hline

0.00000      &( 4 0 )\phantom{*}        &( 3 0 )      &\phantom{-}6.0976        & 59 &30 \\		
0.00000      	 &( 7 0 )\phantom{*}        &( 5 0 )       & \phantom{-}1.0453        & 173 &3               \\
1.87177          &( 3 4 )\phantom{*}        &( 2 3 )      & \phantom{-}1.3725        & 131    &6      \\
3.30431          &( 6 1 )\phantom{*}        &( 4 1 )       & -1.1402        & 149          &6 \\
5.20872          &( 3 1 )\phantom{*}        &( 2 1 )       & \phantom{-}3.8058        & 47   &30\\		
5.20872          &( 2 4 )\phantom{*}        &( 1 3 )       & -3.6090        & 95    &15\\	
8.94828          &( 1 5 )\phantom{*}        &( 0 4 )       & \phantom{-}1.6303        & 110  &12  \\
10.89339         &( 2 1 )\phantom{*}        &( 1 1 )      & -7.3905        & 23  &30\\	
11.30178         &( 4 3 )\phantom{*}        &( 2 3 )       & \phantom{-}1.3725        & 131  &6        \\
12.51983         &{( 7 1 )}*        &( 4 2 )       & -0.8516        & 198      &3\\
13.89789         &( 2 6 )\phantom{*}        &( 0 5 )       & -1.9128        & 179   &3     \\
13.89789         &( 5 0 )\phantom{*}        &( 3 1 )       & \phantom{-}2.0107        & 89  &12    \\		
16.10211         &( 6 2 )\phantom{*}        &( 3 3 )       & \phantom{-}1.9352        & 185    &3  \\
16.10211         &( 3 3 )*        &( 1 3 )       & -1.8401        & 93        &15\\		

19.10661         &( 4 0 )\phantom{*}        &( 2 1 )      & -6.4308        & 53  &18\\	

19.10661         &( 1 2 )\phantom{*}        &( 0 2 )      & \phantom{-}6.9363        & 26 &30 \\	
20.48466         &{( 5 2 )}*        &( 2 3 )       & -1.2610        & 135  &6\\
23.41322         &( 2 3 )\phantom{*}        &( 0 3 )       & -2.6382        & 65  &21      \\		
26.99551         &( 4 2 )\phantom{*}        &( 1 3 )       & -3.6089        & 95    &15  \\		
26.99551         &( 3 1 )\phantom{*}        &( 1 2 )       & \phantom{-}3.8058        & 47  &30     \\		
30.00000     	 &( 5 0 )\phantom{*}        &( 2 2 )       & -1.9913        & 86    &12\\		
30.00000     	 &{( 4 4 )}*        &( 0 5 )       & \phantom{-}2.0924        & 171   &3 \\

\end{tabular}

\end{table}

To quantify the proximity induced SOC in graphene's Dirac bands due to the coupling with the NbSe$_2$ monolayer, we fit the DFT band structures at the Dirac points to a model
Hamiltonian~\citep{Gmitra2015:TMDCgraphene1}. The Hamiltonian $H$ comprises the orbital part $H_{\text{orb}}$ and the spin-orbit part $H_{\text{so}}$. The latter is composed of the 
intrinsic spin-orbit coupling $H_{\text{so,I}}$ and the Rashba coupling $H_{\text{so,R}}$:
\begin{equation}
H(\mathbf{k})=H_{\text{orb}}(\mathbf{k})+H_{\text{so}}=H_{\text{orb}}(\mathbf{k})+
H_{\text{so,I}}+H_{\text{so,R}}.
\label{Eq:Ham}
\end{equation}
The orbital part describes the dispersion of the graphene Dirac cone; it is linearized around the $K$/$K'$-point, therefore $\mathbf{k}$ is the electron wave vector measured from $K$/$K'$. It also includes a staggered potential $\Delta$, taking into account any asymmetrical influence of the NbSe$_2$ substrate on the graphene A- and B-sublattice:
\begin{equation}
H_{\text{orb}}(\mathbf{k})=\hbar v_F (\kappa\sigma_x k_x+\sigma_y k_y)+\Delta \sigma_z.
\end{equation}
Here, $v_F$ is the Fermi velocity of the Dirac electrons and $\sigma_x,\sigma_y$ and $\sigma_z$ are the Pauli matrices operating on the sublattice (A/B) space. The parameter $\kappa$ determines, whether the Hamiltonian describes the band structure near K or K' ($\kappa=1$ for $K$ and $\kappa=-1$ for $K'$).

The intrinsic spin-orbit Hamiltonian
\begin{equation}
H_{\text{so,I}}=\Big[\lambda_{\text{KM}}\sigma_z+\lambda_{\text{VZ}}\sigma_0\Big]\kappa s_z,
\label{Eq:HamKMVZ}
\end{equation}
 and the Rashba spin-orbit Hamiltonian
 \begin{equation}
H_{\text{so,R}}= -\lambda_{\text{R}} \exp(-i\Phi \frac{s_z}{2})
\Big[\kappa\sigma_x s_y-\sigma_y s_x\Big]\exp(i\Phi \frac{s_z}{2}),
\label{Eq:HamR}
\end{equation}
both include spin Pauli matrices $s_x,s_y$ and $s_z$ acting on the spin space; $\lambda_{\text{VZ}}$ and $\lambda_{\text{KM}}$ are the valley-Zeeman~\citep{Gmitra2015:TMDCgraphene1, Wang2015:NC} SOC (sublattice-odd) and the Kane-Mele~\citep{Kane2005:PRL} SOC (sublattice-even) respectively. The Rashba SOC term is defined by two parameters: the magnitude $|\lambda_{\text{R}}|$ and the phase angle $\Phi$. The latter is present in $C_3$ symmetric structures~\citep{Li2019:TwistTB1, David2019:TwistTB2, Naimer2021:paper1} and rotates the spin texture about the $z$-axis, adding a radial component to the Rashba field.

We only construct heterostructures with angles between 0\degree~and 30\degree. The parameters for all other twist angles can be obtained by the following symmetry rules:
\begin{align}
\lambda_{\text{VZ}}(-\Theta)&=\lambda_{\text{VZ}}(\Theta) \\
|\lambda_{\text{R}}(-\Theta)|&=|\lambda_{\text{R}}(\Theta)|\\
\Phi(-\Theta)&=-\Phi(\Theta)\\
\Delta(-\Theta)&=\Delta(\Theta)\\
\lambda_{\text{VZ}}(\Theta+60\degree)&=-\lambda_{\text{VZ}}(\Theta)\\
|\lambda_{\text{R}}(\Theta+60\degree)|&=|\lambda_{\text{R}}(\Theta)|\\
\Phi(\Theta+60\degree)&=\Phi(\Theta)\\
\Delta(\Theta+60\degree)&=-\Delta(\Theta).
\end{align}

We note that NbSe$_2$ exhibits charge density wave (CDW)~\cite{Ugeda2016:NP,Xi2015:NN,Lian2018:NL}.
In App.~\ref{App:relax}, we discuss a typical $3\times3$ CDW in NbSe$_2$ with the same atomic rearrangement in the heterostructure with graphene. Since the CDW does not significantly influence the proximity SOC in graphene, in the main text we perform simulations on unrelaxed structures to facilitate systematic comparison between different twist angles. 

For DFT calculations of heterostructures, the varying band offsets induced by strain (see Sec.~\ref{Sec:bandoffsets}) can be relevant for the extracted SOC parameters. In Refs.~\cite{Naimer2021:paper1,Naimer2023:paper2} we used electric fields to correct for the strain-induced band offsets. However, in this paper we use only supercells with built-in strain $\epsilon<5\%$ for the determination of the SOC parameters. Therefore, we find a much lower variance in band offsets and hence rather follow the approach of Refs.~\cite{Lee22:PRB:radialrashba3,Zollner2023:PRB} and deem these corrections unnecessary.  In App.~\ref{App:EVSSOC} we explore how a transverse electric field can influence the SOC parameters and orbital contributions for one particular example. Our findings indicate that the electric field's main effect is a change of the Rashba phase angle $\Phi$.

\section{Energetic alignments}
\label{Sec:bandoffsets}
\begin{figure}
    \includegraphics[width=.99\linewidth]{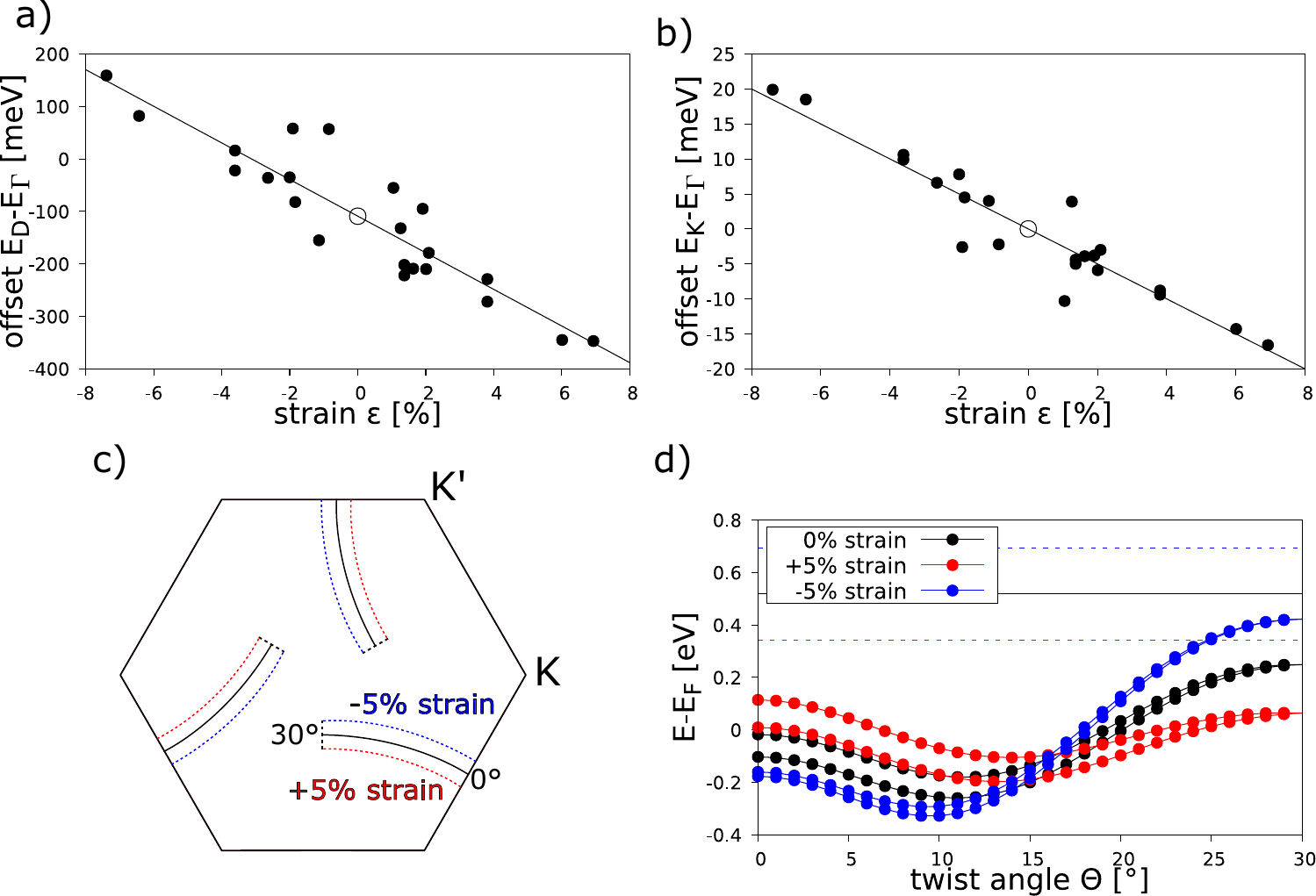}
    \caption{Influence of the strain $\epsilon$ in graphene on the energetic alignments in the heterostructure. The band offsets of both the graphene Dirac cone (a) and the NbSe$_2$ $K$ band (b) with respect to the NbSe$_2$ $\Gamma$ band are fitted linearly (solid black line). Each of the solid circles represents one of the heterostructures in Tab.~\ref{Tab:structures}. Subfig. c) shows the first Brillouin zone of NbSe$_2$ with the ${k}$ points to which the Dirac cone couples by generalized Umklapp processes for twist angles between 0$^\circ$ and 30$^\circ$. We show the paths for three different strains (blue, black, red). In d) the energies of the states close to the Dirac cone along the three paths (blue, black, red) in c) are shown. For reference we additionally indicate an estimation of the position of the Dirac cone (using the fit from a)) as vertical lines for different strains (again blue, black red). Note that for all cases the applied strain is still always applied to graphene.
    \label{Fig:strainVSoffset}}
\end{figure}

\begin{figure}
    \includegraphics[width=.99\linewidth]{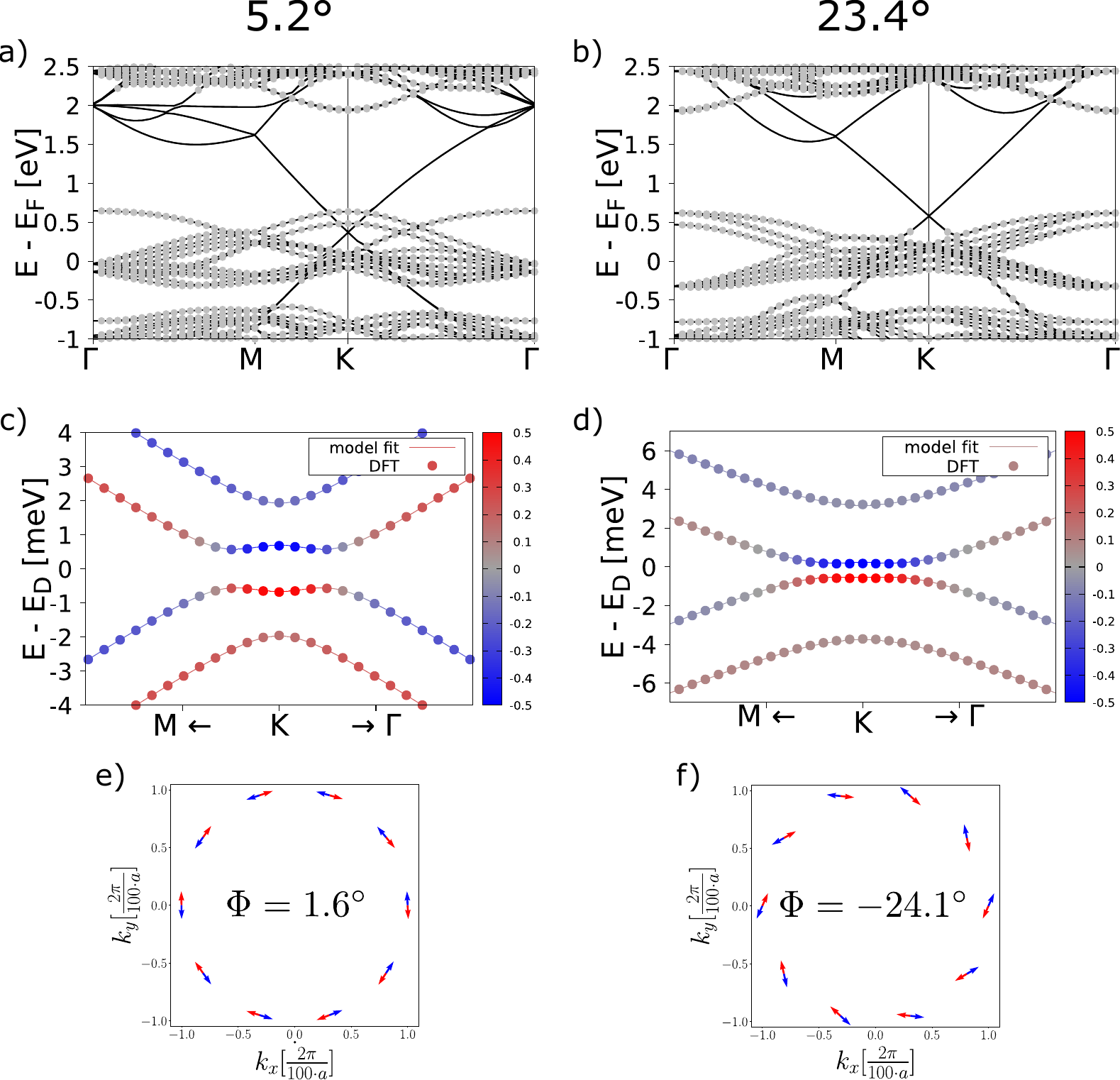}
    \caption{Band structure of two examples of graphene/NbSe$_2$ heterostructures ($\Theta=5.2^\circ$ and $\Theta=23.4^\circ$). a) and b) show the calculated band structure along high symmetry points. Grey bands originate from the NbSe$_2$, while the black lines originate from graphene. c) and d) show a zoom to the graphene Dirac cone, with color coded spin-z expectation value. The dots represent DFT data points and the solid line shows the fit by the model Hamiltonian $H(\mathbf{k})$, see Eq.~\ref{Eq:Ham}. e) and f) show the in-plane spin structure around a circular $k$-path around the Dirac cone (red arrows: energetically lower valence band, blue arrows: energetically higher valence band).\label{Fig:BSglobal}}
\end{figure}

We calculate the band structures of our supercell graphene/NbSe$_2$ heterostructures using DFT. Two examples can be seen in Fig.~\ref{Fig:BSglobal} a) and b). In contrast to semiconducting graphene/TMDC heterostructures (Ref.~\cite{Naimer2021:paper1,Gmitra2015:TMDCgraphene1,Gmitra2016:TMDCgraphene2,FariaJunior2023:2DM:TMDCGrgfactors}) the graphene Dirac cone (solid black lines) lies buried within the NbSe$_2$ bands (grey bands) for most cases. However, in some cases the Dirac cone lies above these bands. To better describe the energetic alignments we define the band offsets as follows: $E_{\Gamma}$ and $E_K$ are the band edges of the NbSe$_2$ states near the Fermi level at $\Gamma$ and $K$ respectively. $E_D$ is the energy level of the Dirac cone. In Fig.~\ref{Fig:strainVSoffset} a) we show the (approximately linear\cite{Grassano2020:PRB:graphenestrainworkfunction}) relation between the strain $\epsilon$ in graphene and the band offset $E_D-E_{\Gamma}$. If this band offset is positive, the Dirac cone lies above the NbSe$_2$ bands. By applying a linear fit we can extract a zero-strain band offset $E_D-E_{\Gamma}=-109$~meV. Defining the band offset as $E_D-E_K$ would yield very similar results.
Nevertheless, we need to distinguish between $E_{\Gamma}$ and $E_K$ because the NbSe$_2$ band structure also changes by the proximity of the graphene. In fact, we see a similar behaviour (linear with strain $\epsilon$, see Fig.~\ref{Fig:strainVSoffset} b)) of the internal band offset $E_K-E_{\Gamma}$. Depending on $\epsilon$, this band offset can also be negative or positive. This can be interpreted in the following way: Compressing the graphene ($\epsilon <0$) creates a more dense barrier of graphene $p_z$ orbitals pushing the (out-of-plane orbital dominated) NbSe$_2$ $\Gamma$ bands further down in energy than the (more in-plane) $K$ bands.

Although for most of the heterostructure supercells the Dirac cone lies within the NbSe$_2$ bands, this does not necessarily mean that the Dirac cone hybridizes with NbSe$_2$ bands in a way that makes it impossible to describe the Dirac cone with our model Hamiltonian. The reason for this is that the NbSe$_2$ bands, which are energetically close to the Dirac cone, hardly interact with it. Following the theory of generalized Umklapp processes~\cite{Koshino2015:TwistTBBasic} the bands interacting the most with the Dirac cone are energetically well separated from it. This is illustrated in Fig.~\ref{Fig:strainVSoffset} c) and d): Depending on strain and twist angle, the Dirac cone interacts with a different $k$ point in the first Brillouin zone of the primitive NbSe$_2$ (shown in Fig.~\ref{Fig:strainVSoffset} c)). We are now concerned with the NbSe$_2$ bands at this $k$-point, which are energetically closest to the Dirac point; their energies are shown in Fig.~\ref{Fig:strainVSoffset} d). The Dirac cone energy resides between the blue and red dotted line (depending on $\epsilon$) and is always at least 200~meV away from the relevant NbSe$_2$ state. Another factor changing by the same mechanism is the orbital composition of the relevant NbSe$_2$ states. This is shown in Fig.~\ref{Fig:SOC} d) and discussed in Sec.~\ref{Sec:SOC}.

While we showed why there are only crossings (instead of anti-crossings) between the Dirac cone and any nearby NbSe$_2$ bands, there is still the issue of distinguishing proximitized graphene transport properties from NbSe$_2$ transport properties in experiments. For example, in a charge-to-spin conversion experiment, this distinction can be made by shifting the Fermi level by gating. A sign change of the signal, which is expected for the Dirac cone (see Fig.~\ref{Fig:UREE} a)), is a clear indicator that the signal is coming from the proximitized graphene rather than the NbSe$_2$.

\section{Proximity SOC}
\label{Sec:SOC}
\begin{figure}
    \includegraphics[width=.99\linewidth]{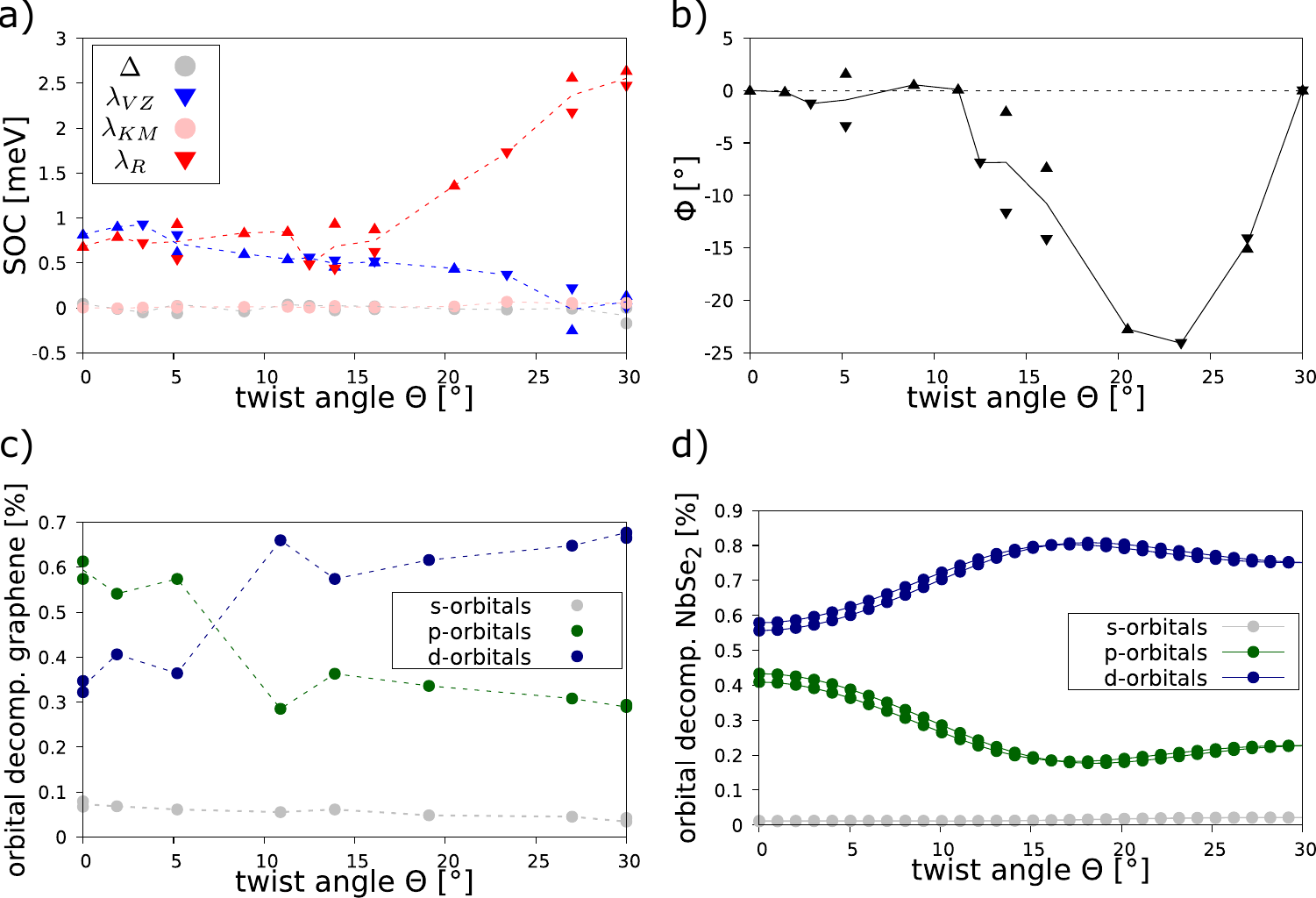}
    \caption{Twist-angle dependencies of a) the staggered potential $\Delta$ and the SOC parameters $\lambda_R, \lambda_{VZ},\lambda_{KM}$,
    b) dependence of $\Phi$.
    Upward/downward pointing triangles indicate data points with tensile ($\epsilon>0$)/compressive ($\epsilon<0$) strains.
    The dotted lines in a) and solid line in b) are a mere guide to the eyes. For data points with a Dirac cone backfolded to $\Gamma$, the sign of $\lambda_{VZ}$ cannot be determined and is assumed to be positive in accordance with the other supercells. c) shows what NbSe$_2$ orbitals appear in the proximitized Dirac cone of heterostructures with different twist angles. d) shows the orbital decomposition of the nearest NbSe$_2$ bands interacting with the Dirac cone by generalized Umklapp processes (see Fig.~\ref{Fig:strainVSoffset} d)).\label{Fig:SOC}}
\end{figure}

Fitting the low energy Dirac bands to the model Hamiltonian (see Eq.~\ref{Eq:Ham}) gives the SOC parameters listed in Tab.~\ref{Tab:results} and shown in Fig.~\ref{Fig:SOC}. As for other unrelaxed TMDC/graphene heterostructures (with semiconducting TMDCs, see Ref.~\cite{Naimer2021:paper1}), the Kane-Mele SOC $\lambda_{KM}$ and the staggered potential $\Delta$ are negligibly small. Starting at $\Theta=0^\circ$ twist angle, both Rashba SOC ($\lambda_R$) and valley-Zeeman SOC ($\lambda_{VZ}$) have very similar size of about 0.5~meV to 1~meV. They stay at this level, until at $\Theta\approx 15^\circ$ $\lambda_R$ starts to increase up to a value of $\lambda_R\approx2.5$~meV at $\Theta=30^\circ$. Simultaneously $\lambda_{VZ}$ decreases and vanishes at 
 $\Theta=30^\circ$, which is a feature also seen in TMDC/graphene heterostructures with semiconducting TMDCs. Moreover, at roughly the same mark of $\Theta\approx 15^\circ$, the Rashba phase angle $\Phi$ increases as well, peaking at about $\Theta=22^\circ$ with a value of $\Phi=-25^\circ$. After this peak, $\Phi$ rapidly decreases to $\Phi=0^\circ$ at $\Theta=30^\circ$, which is demanded by symmetry. 

We argue that the main contribution of the proximity effect comes from the bands depicted in Fig.~\ref{Fig:strainVSoffset} d). This seems plausible since they are the energetically closest bands coupled to the Dirac cone through generalized Umklapp processes (see Ref.~\cite{Koshino2015:TwistTBBasic}). Additionally, an orbital analysis shows that the orbital composition of these bands (Fig.~\ref{Fig:SOC} d)) compares well to the orbital composition of the NbSe$_2$ orbitals found in the proximitized graphene's Dirac cone (Fig.~\ref{Fig:SOC} c).
Note that although both in Fig.~\ref{Fig:SOC} c) and Fig.~\ref{Fig:SOC} d) the twist-angle dependence is depicted, the data points in Fig.~\ref{Fig:SOC} c) come from the heterostructure calculations and the data points in Fig.~\ref{Fig:SOC} d) come from calculations of a NbSe$_2$ monolayer along the black path depicted in Fig.~\ref{Fig:strainVSoffset} c). 
The twist-angle dependence takes the following form in both cases: While s-orbital contribution is negligible, p- and d-orbitals contribute similarly, with p-orbitals more dominant at twist angles near $0^\circ$. There is one notable difference between the orbital composition of NbSe$_2$ along the twist angle path and the orbital composition found in the actual heterostructures. Namely, that the ratio of p- and d-orbitals is slightly shifted in favour of the p-orbitals for the latter case for all twist angles. This can be attributed to the fact that the NbSe$_2$ d-orbitals come exclusively from the Nb atoms and are therefore located farther away from the graphene.

Assuming these bands are responsible for the proximity SOC in the Dirac cone leads us to a plausible explanation for the twist-angle dependence seen in Fig.~\ref{Fig:SOC} a): For moderate strain (see black curve in Fig.~\ref{Fig:strainVSoffset} d)) the relevant contributing bands move towards the Dirac cone in energy, when going from $\Theta=0^\circ$ to $\Theta=30^\circ$. At the same time, the splitting between the two spin-split subbands decreases and finally vanishes for $\Theta=30^\circ$. The general increase of the proximity SOC, while twisting from $\Theta=0^\circ$ to $\Theta=30^\circ$ can be seen as a consequence of the first point; if the contributing states are closer in energy, their influence and therefore the proximity SOC is expected to grow. Nevertheless, the decrease of the spin-splitting causes a decrease of the valley Zeeman SOC. This is to be expected, as the proximity valley-Zeeman SOC is mainly driven by a spin split in the NbSe$_2$ substrate bands. However, this explanation is not complete, since it does not fully account for a possible change of orbital overlap between the Dirac cone and the NbSe$_2$ bands. Especially, the large increase in Rashba SOC exceeds the expectation one might have from this simple picture. Also, the increase of the Rashba phase angle $\Phi$, which arises from a rather complex interference of phases, cannot be explained by such a handwaving argument. This underlines the need for the full DFT treatments, as done in this paper.

 \begin{table}[htb]
   \caption{Parameters extracted from the band structure calculations. For all angles, we list the extracted model Hamiltonian (Eq.~\eqref{Eq:Ham}) parameters and the band offset $\Delta E=E_D-E_{\Gamma}$  of the Dirac cone with respect to the NbSe$_2$ $\Gamma$ band (see Fig.~\ref{Fig:strainVSoffset} a). The parameters are staggered potential $\Delta$, Kane-Mele SOC $\lambda_{KM}$, valley-Zeeman SOC $\lambda_{VZ}$, magnitude of the Rashba SOC $|\lambda_{\text{R}}|$ and Rashba angle $\Phi$.  For some of the supercells the Dirac cone is folded back to $\Gamma$ ($n-m=3\cdot k, k\in \mathbb{Z}$). As a consequence, the sign of the $\lambda_{VZ}$ cannot be determined unambiguously and is presented with a $\pm$.}
    \label{Tab:results} 
    \begin{ruledtabular}
    \begin{tabular}{cc|cccccc}

$\Theta[\degree]$&$\epsilon$&$\Phi$&$\Delta$ & $\lambda_{KM}$ & $\lambda_{VZ}$ & $|\lambda_{\text{R}}|$& $\Delta E$ \\

&[\%]&[\degree]&[meV] & [meV] & [meV] & [meV]& [eV] \\
\hline
 0.0& \phantom{-}6.10 & \phantom{-}0 & \phantom{-}0.081 & -0.001 & 0.913 & 0.846  & -0.345   \\
 0.0& \phantom{-}1.05 & \phantom{-}0 & \phantom{-}0.045 & \phantom{-}0.002 & \phantom{-}0.817 & 0.681  &  -0.055  \\
 1.9& \phantom{-}1.37 & \phantom{-}0 & \phantom{-}0.013 & -0.005 & \phantom{-}0.902 & 0.791  & -0.202   \\
 3.3& -1.14 & -1 & \phantom{-}0.049 & \phantom{-}0.003 & \phantom{-}0.925 & 0.716  & -0.155   \\
 5.2& -3.61 & -3 &\phantom{-}0.021 & \phantom{-}0.002 & \phantom{-}0.807 & 0.538  & -0.022   \\
 5.2& \phantom{-}3.81 & \phantom{-}2 & \phantom{-}0.060 & \phantom{-}0.019 & \phantom{-}0.619 & 0.934  &  -0.272  \\
 8.9& \phantom{-}1.63 & \phantom{-}1 & \phantom{-}0.039 & \phantom{-}0.010 & \phantom{-}0.601 & 0.833  &   -0.209  \\
 10.9& -7.39 & -19 & -0.016 & \phantom{-}0.003 & \phantom{-}0.452 & 0.225  &  \phantom{-}0.159  \\
 11.3& \phantom{-}1.37 & \phantom{-}0 & \phantom{-}0.036 & \phantom{-}0.012 & \phantom{-}0.542 & 0.848  &  -0.222  \\
 12.5& -0.85 & -7 & \phantom{-}0.023 & \phantom{-}0.002 & $\pm$0.555 & 0.481  &  \phantom{-}0.057  \\
 13.9 & -1.91 & -12 & \phantom{-}0.019 & \phantom{-}0.003 & \phantom{-}0.524 & 0.433  &   \phantom{-}0.058  \\
 13.9& \phantom{-}2.01 & -2 & \phantom{-}0.028 & \phantom{-}0.021 & \phantom{-}0.460 & 0.937  &   -0.210  \\
 16.1&\phantom{-}1.94 & -7 & \phantom{-}0.017 & \phantom{-}0.010 & $\pm$0.507 & 0.877  &   -0.095  \\
 16.1& -1.84 & -14 & \phantom{-}0.015 & \phantom{-}0.004 & \phantom{-}0.516 & 0.619  &   -0.082  \\
 19.1& \phantom{-}6.93 & -10 & -0.002 & \phantom{-}0.035 & -0.438 & 1.973  &    -0.347 \\
 19.1& -6.43 & -33 & \phantom{-}0.009 & \phantom{-}0.006 & \phantom{-}0.315 & 0.467  &    \phantom{-}0.082 \\
 20.5& -1.26 & -23 & -0.014 & \phantom{-}0.015 & $\pm$0.438 & 1.361  &    -0.132 \\
 23.4& -2.64 & -24 & \phantom{-}0.018 & \phantom{-}0.067 & \phantom{-}0.366 & 1.726  &  -0.036  \\
 27.0& -3.61 & -14 & \phantom{-}0.009 & \phantom{-}0.043 & \phantom{-}0.214 & 2.17  &     \phantom{-}0.016 \\
 27.0& \phantom{-}3.81 & -15 & -0.019 & \phantom{-}0.054 & -0.249 & 2.562  &    -0.229 \\
 30.0& -2.00 & \phantom{-}0 & \phantom{-}0.000 & \phantom{-}0.053 & \phantom{-}0.000 & 2.468  &  -0.035   \\
 30.0& \phantom{-}2.09 & \phantom{-}0 & -0.173 & \phantom{-}0.052 & $\pm$0.134 & 2.638  &   -0.179 \\
    \end{tabular}
    \end{ruledtabular}
    \end{table}
\section{Charge-to-spin conversion efficiencies}
\label{Sec:UREE}
The Rashba SOC (and the resulting in-plane spin structure) in graphene based heterostructures enables the Rashba Edelstein effect~\cite{Ghiasi2017:NL,Offidani2017:PRL,Drydal2014:PRB,Edelstein1990:SSC}  (REE) and unconventional Rashba Edelstein effect~\cite{Ingla_Aynes2022:expradialrashba2,Peterfalvi2022:radialrashba,Veneri22:PRB:radialrashba2,Lee22:PRB:radialrashba3,Camosi_2022:2DM:expradialrashba3,Ontonso2023:PRA:expradialrashba4,yang2024:arxiv:GRTMDCradialexp} (UREE) as efficient ways for charge-to-spin conversion. In the REE, the accumulated spins are perpendicular to the charge current, while they are collinear to it in the UREE.
We analyze the potential of graphene proximitized by NbSe$_2$ for this kind of charge-to-spin conversion. By using the Kubo formula ~\cite{Kubo1956:CJP,Kubo1957:JPSJ} in the Smrcka-Streda ~\cite{Smrcka1977:JPCSSP,Crepieux2001:PRB,Bonbien2020:PRB} formulation, we calculate the REE and UREE.

We assume weak disorder scattering describe by phenomenological parameter $\gamma$~\cite{Freimuth2014:PRB,Zelezny2017:PRL,Bonbien2020:PRB,Lee22:PRB:radialrashba3}.
The change $\delta O$ in the observable $O$ is given by
\begin{equation}
    \delta O =\frac{E}{4\pi ^2} \int d^2 \mathbf{k}[ \chi^{surf}_O(\mathbf{k})+\chi^{sea}_O(\mathbf{k})],
\end{equation}
with the electric field strength $E$ and Fermi sea and Fermi surface susceptibilities $\chi^{surf}_O$ and $\chi^{sea}_O$. Writing the formula in this form assumes that the electric field points only in one direction (in our case $x$-direction). In a more general form the susceptibilities need an additional index, effectively forming a susceptibility tensor.

The integral is performed around $K$ and $K'$ points ensuring to cover all the relevant states (151$\times$151 square grid with a side length of $\Delta k=0.03\frac{1}{v_F\hbar}\approx2.5\cdot10^{-3}\frac{2\pi}{a}$). 
The integrands $\chi^{surf}_O$ and $\chi^{sea}_O$ are given by:
\begin{widetext}
\begin{equation}
    \chi^{surf}_O(\mathbf{k})=\frac{\hbar}{\pi}\gamma^2\sum_{n,m}\frac{\mathrm{Re} \big( \bra{n\mathbf{k}}\hat{O}\ket{m\mathbf{k}}\bra{m\mathbf{k}}-e\hat{v}_x\ket{n\mathbf{k}}
    \big)}{[(\epsilon_{n\mathbf{k}}-E_F)^2)+\gamma^2]\cdot[(\epsilon_{m\mathbf{k}}-E_F)^2)+\gamma^2]}\label{Eq:surf}
\end{equation}
and
\begin{equation}
    \chi^{sea}_O(\mathbf{k})=\hbar\sum_{n\neq m}(f_{n,\mathbf{k}}-f_{m,\mathbf{k}})\frac{\mathrm{Im}\big(\bra{n\mathbf{k}}\hat{O}\ket{m\mathbf{k}}\bra{m\mathbf{k}}-e\hat{v}_x\ket{n\mathbf{k}}\big)}
    {(\epsilon_{n\mathbf{k}}-\epsilon_{m\mathbf{k}})^2}.\label{Eq:sea}
\end{equation}
\end{widetext}
Here, $\epsilon_{n\mathbf{k}}$ and $\ket{n\mathbf{k}}$ are the eigenenergy and eigenstate of band $n$ as represented by the model Hamiltonian Eq.~(\ref{Eq:Ham}) at $\mathbf{k}$. We set the broadening $\gamma$ of the states to $\gamma=0.1$~meV and the electronic temperature of the Fermi functions $f_{n\mathbf{k}}$ to $k_BT=0.01$~meV.
In our setup, the perturbation is always given by an electric field in the $x$-direction, hence we used $-e\hat{v}_x$ in the second matrix element for $\chi^{surf}_O$ and $\chi^{sea}_O$.
The physical observables are either $O=s_{x/y}$ for spin accumulation in $x$- or $y$-direction, with
\begin{equation}
   \hat{O}=\hat{s}_{x/y}=\frac{\hbar}{2}\sigma_{x/y},
\end{equation}
or $O=j_x$ for the charge density response, with 
\begin{equation}
    \hat{O}=\hat{j}_x=-e\hat{v}_x=-e\frac{1}{\hbar}\frac{\partial H}{\partial k_x}=-e  v_F \kappa \sigma_{x}.
\end{equation} 
As in Ref.~\cite{Lee22:PRB:radialrashba3}, we calculate the efficiency of the REE (or UREE) as the spin-y (or spin-x) accumulation normalized by the  charge density response, so:
\begin{equation}
    \alpha_{\mathrm{REE}}=\frac{v_Fe}{\hbar}\frac{\delta s_y}{\delta j_x} \mathrm{\phantom{....}and\phantom{....}}
    \alpha_{\mathrm{UREE}}=\frac{v_Fe}{\hbar}\frac{\delta s_x}{\delta j_x} .
\end{equation}

\begin{figure*}
    \includegraphics[width=.99\linewidth]{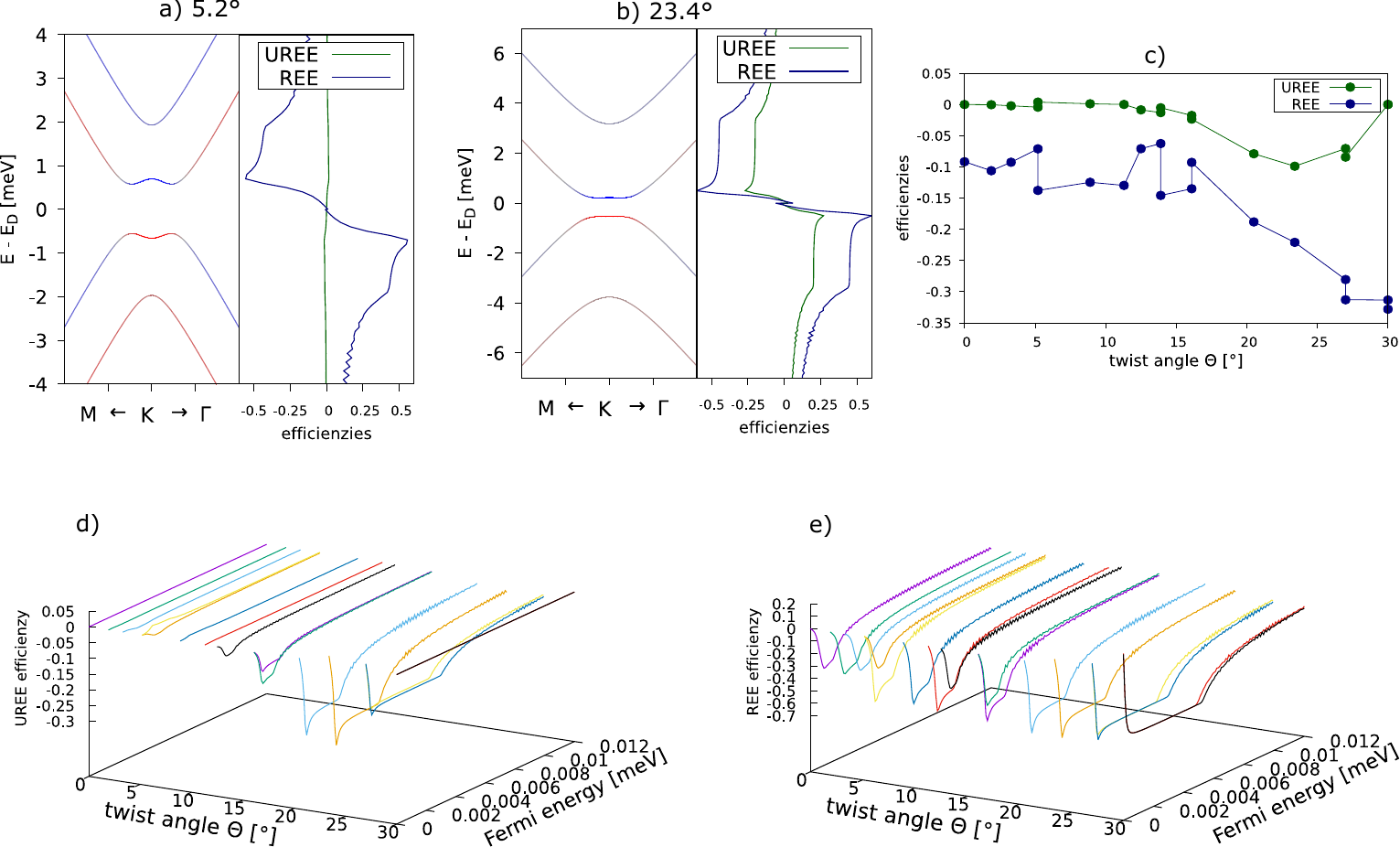}
    \caption{Calculated charge-to-spin conversion efficiencies in graphene on NbSe$_2$. a)~band structure in vicinity of the Dirac cone with spin-z color coded and UREE and REE dependencies for the twist angle $\Theta=5.2\degree$; b)~similar as in a) but for $\Theta=23.4\degree$. 
    c) conversion efficiencies as a function of the twist angle averaged over 12~meV ($E_F=0$ to $E_F=12~$meV).
    d) shows the Fermi-energy dependence of UREE for all twist angles with strain $\epsilon<5\%$.  e) shows the same as d), but for REE.
    \label{Fig:UREE}}
\end{figure*}

Fig.~\ref{Fig:UREE} a) and b) show the Fermi-energy dependent charge-to-spin efficiencies (REE and UREE) for two twist angles. Since for both cases Rashba SOC is dominating, the pattern (in accordance with Ref.~\cite{Lee22:PRB:radialrashba3}) is the following: First, we note that as reported by other groups~\cite{Lee22:PRB:radialrashba3,Drydal2014:PRB} both REE and UREE are anti-symmetric with respect to the Fermi level position. This means, that if the Fermi level lies in the Dirac cone valence bands, the sign of the charge-to-spin conversion is opposite compared to the case where the Fermi level lies in the conduction bands (see Fig.~\ref{Fig:UREE} a) and b)). Hence, the charge-to-spin conversion in the middle of the small band gap is zero. However, as soon as the Fermi level coincides with one of the inner bands, there is a peak in the charge-to-spin conversion. After this initial peak the efficiencies plateau, only to decrease again after the Fermi level touches the second band. The reason for the initial peak is the flatness of the bands at this energy, which comes from the presence of the valley-Zeeman SOC. In the plateau region the slope of the bands is constant and although $\delta s_{x/y}$ increase (with increasing radius of the relevant contributing band), since $\delta j_x$ increases at the same rate, the total charge-to-spin efficiencies stay constant. As soon as the second band starts contributing, the total efficiencies decay, because contributions from the two bands are of opposite sign. From this point on, the finite difference in contribution from the two bands stops the growth of $\delta s_{x/y}$. Therefore the charge-to-spin conversion vanishes in the limit of $E_F\rightarrow \pm\infty$, since $\delta j_x$ keeps increasing. This pattern is the same for both REE and UREE. 
While a) presents a case where the Rashba phase angle $\Phi$ is very small, b) shows a case where $\Phi=-24^\circ$. Therefore there is almost no sign of UREE in a), while in b) UREE and REE are comparable in size.

Fig.~\ref{Fig:UREE} d) and e) show UREE and REE respectively for all different investigated angles with strains $|\epsilon |<5\%$. From these plots, one can already see the two main observations:
\begin{enumerate}
\item UREE peaks at $\Theta\approx22^\circ$, the same twist angle that $\Phi$ peaks at. This is very natural, since $\Phi$ enables the radial in-plane spin structure and therefore UREE.
\item Although the maximum value for the REE does not change drastically throughout all twist angles, the length of the plateau between the two conduction bands grows after $\Theta=15^\circ$ and is largest at $\Theta =30^\circ$. This will lead to larger charge-to-spin yields, since in experiment the Fermi level cannot be fine tuned with the precision needed to exactly meet the peak.
\end{enumerate}    
We visualize both points by plotting the (U)REE efficiencies averaged over 12~meV ($E_F=0$ to $E_F=12~$meV) in Fig.~\ref{Fig:UREE} c). Using the peak efficiency instead of the average efficiency will give different results. We show those along with the results of an alternative (U)REE calculation (using Ref.~\cite{Veneri22:PRB:radialrashba2}) in App.~\ref{App:UREEsupp}.

\section{Summary}
We performed DFT calculations on several twisted graphene/NbSe$_2$ supercells and extracted SOC parameters using a model Hamiltonian. Based on those fittings we additionally performed Kubo formula calculations giving us spin responses to an electric field and therefore (collinear and perpendicular) charge-to-spin conversion efficiencies. Since the heterostructure supercells have different twist angles $\Theta$, we can establish a twist-angle dependence for all SOC parameters and charge-to-spin conversion efficiencies.
We find Rashba SOC $\lambda_R$ to increase by a factor of 3 going from $\Theta=0^\circ$ to $\Theta=30^\circ$. Furthermore, the Rashba phase angle $\Phi$ peaks at $\Theta\approx 23^\circ$. Consequently the perpendicular (REE) and collinear (UREE) charge-to-spin conversion efficiencies are also maximal at $\Theta=30^\circ$ and $\Theta\approx23^\circ$ respectively.

Additional investigations on the effect of external electric field and relaxation were performed on sample heterostructures. They indicate that the main effect of the electric field is an increase of the Rashba phase angle $\Phi$, while the main effect of relaxation is the change of interlayer distance and the resulting increase in general SOC strength. The typical $3\times3$ rearrangement (linked to a CDW state) is observed, but does not change the proximity SOC significantly.

\acknowledgments All authors acknowledge support by the FLAG ERA JTC 2021 project 2DSOTECH. T.~N. and J.~F. were also supported by the European Union Horizon 2020 Research and Innovation Program 881603 (Graphene Flagship). M.G.~acknowledges additional financial support provided by the Slovak Research and Development Agency provided under Contract No. APVV-SK-CZ-RD-21-0114 and Slovak Academy of Sciences project IMPULZ IM-2021-42. The authors gratefully acknowledge the Gauss Centre for Supercomputing e.V. (www.gauss-centre.eu) for funding this project by providing computing time on the GCS Supercomputer SUPERMUC-NG at Leibniz Supercomputing Centre (www.lrz.de).

\appendix

\section{Effects of electric field}
\label{App:EVSSOC}
\begin{figure}
    \includegraphics[width=.99\linewidth]{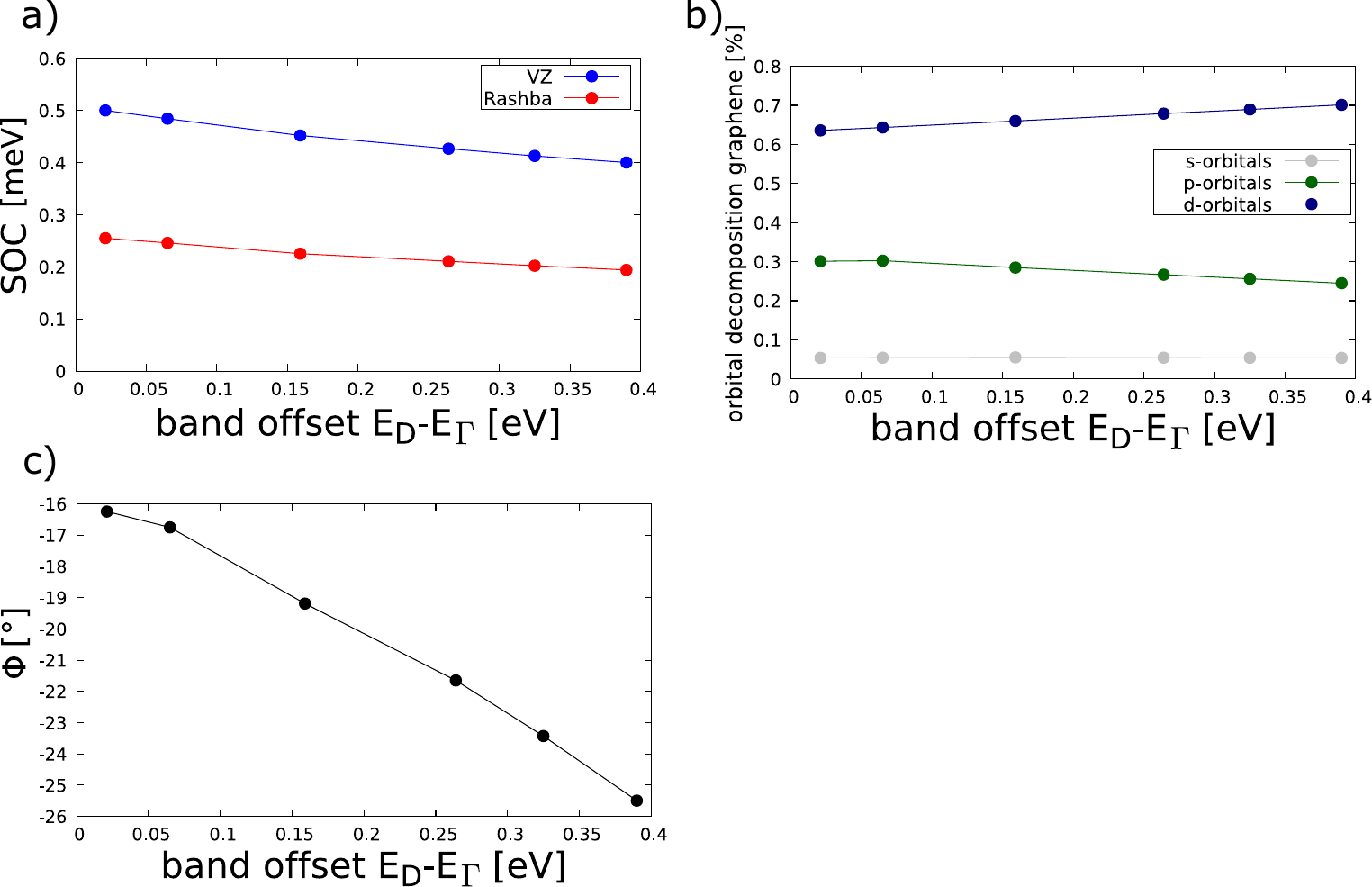}
    \caption{Effect of a transverse electric field on the proximity SOC of the 10.9$^\circ$ supercell. The band offset (determined by the electric field) is plotted against the valley-Zeeman and Rashba SOC (a) as well as the Rashba phase angle $\Phi$ (c). b) shows the change of the NbSe$_2$ orbitals which are contributing to the Dirac cone.\label{Fig:EVSSOC}}
\end{figure}

The exact position of the Dirac cone with respect to the NbSe$_2$ bands can have a very relevant influence on the proximity SOC. In order to study the band offset effect (as defined in Sec.~\ref{Sec:bandoffsets}), we apply a transverse electrical field. The resulting SOC parameters are shown in Fig.~\ref{Fig:EVSSOC} a) and c). While the Dirac cone is moved from very close vicinity to the NbSe$_2$ bands ($E_D-E_{\Gamma} = 0.021$~eV) to well within the band gap ($E_D-E_{\Gamma} = 0.391$~eV), the Rashba SOC $\lambda_R$ and valley-Zeeman SOC $\lambda_{VZ}$ are decaying slightly. This decay seems to be rather weak since the lower bound for the band offset is almost zero (Dirac cone nearly touching the NbSe$_2$ states). However, this is more understandable considering that those very close NbSe$_2$ states do not interact with the Dirac cone. The nearest states actually interacting with the Dirac cone are 800~meV lower in energy. Contrary to the rather small change of $\lambda_R$ and $\lambda_{VZ}$, the Rashba phase angle $\Phi$  nearly doubles in magnitude, while the Dirac cone is being shifted further into the band gap. Fig.~\ref{Fig:EVSSOC} b) shows which NbSe$_2$ bands contribute to the Dirac cone. Except for a slight shift towards $d$-orbitals, the orbital composition stays roughly constant.

\section{Effects of relaxation}
\label{App:relax}
\begin{figure}
    \includegraphics[width=.99\linewidth]{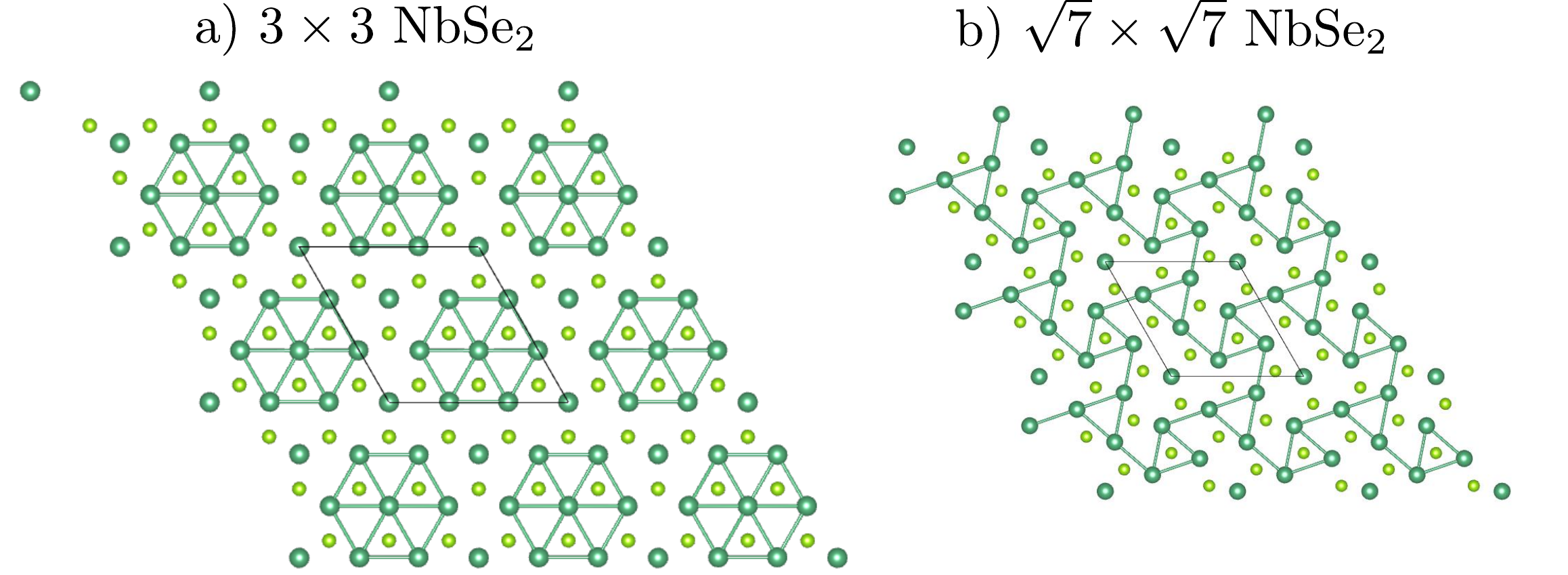}
    \caption{Relaxed structures for two examples of graphene/NbSe$_2$ heterostructures. Nb atoms are dark green, Se atoms are light green, carbon atoms are omitted. To show the subtle atomic reconstructions in the two supercells, we only plot bonds between Nb atoms, if they are less than $3.48~$\AA~apart. The black lines indicate the ($3\times3$ or $\sqrt{7}\times\sqrt{7}$) supercell. More details and SOC parameters are listed in Tab.~\ref{Tab:relax}. \label{Fig:relax}}
\end{figure}

All results in the main paper are concerned with idealized structures with a fixed interlayer distance and unrelaxed structures of both graphene and NbSe$_2$. The results of Ref.~\cite{Naimer2021:paper1} indicate that the relaxation of graphene/transition-metal dichalcogenide heterostructures has little direct effect on the proximity SOC. Instead it enhances the unwanted effects (such as an increase of the staggered potential $\Delta$) of the strain necessary to construct the commensurate supercell. However, NbSe$_2$ is known to exhibit a periodic lattice distortion with charge density wave for 3$\times 3$ supercells~\cite{Ugeda2016:NP,Xi2015:NN,Lian2018:NL}, where the atoms rearrange in a triangular or possibly other~\cite{Guster2019:NL} structure. Since this might introduce relevant changes in electronic structure, we performed additional relaxation calculations on $3\times 3$ and larger $\sqrt{7}\times\sqrt{7}$ heterostructures. The results of these relaxations are shown in Fig.~\ref{Fig:relax}, where we plot the most prevalent Nb-Nb bonds (bond lengths must be bigger than the unrelaxed Nb-Nb distance of 3.48~\AA). By doing this, we show the rearrangement of the 3$\times3$ supercell into filled hexagons.
It is the same pattern, which arises, if we perform the relaxation calculations without the graphene layer. This shows that (at least regarding the rearrangement) the CDW phase is unchanged by the nearby layer of graphene. The other supercell for comparison shows another pattern, which is most likely caused by the nearby graphene layer. Its maximal difference in Nb-Nb bond length is much smaller (about $7.66~$m\AA) than the one of the $3\times3$ supercell (about $72.44~$m\AA).

Naturally, the question arises: Does this rearranging change the proximity SOC in either of the two heterostructures? For both heterostructures $\lambda_R$ and $\lambda_{VZ}$ decrease by about 35\% (see Tab.~\ref{Tab:relax}). However, this can be attributed to the change of the interlayer distance between graphene and NbSe$_2$, which we apparently slightly underestimated in the calculations of the main paper. The equilibrium interlayer distance as determined by the relaxation calculations is about 3.5\% larger than the one we assumed. According to the interlayer distance study presented in Ref.~\cite{Gmitra2016:TMDCgraphene2} the decrease in SOC is roughly what can be expected for such an increased interlayer distance. 
Additionally, as expected, through the relaxation there is an increase in the parameters $\lambda_{KM}$ and $\Delta$, which are suppressed in the idealized structures.
 Based on our limited relaxation calculations, we estimate that the formation of the rearranged triangular pattern typical for CDW can largely be considered not relevant for the proximity SOC effects. 
\begin{table}[]
    \centering
    \begin{tabular}{cc|cc|c|ccccc}
         $(n,m)$& $(n',m')$ & $\Theta$& $\epsilon$& relaxed & $\Phi$&$\Delta$ & $\lambda_{KM}$ & $\lambda_{VZ}$ & $|\lambda_{\text{R}}|$ \\
               &       & [$^\circ$] &[$\%]$& &  [$^\circ$]     & [meV]    &[meV]     &[meV]     &[meV]     \\
                \hline
              (4,0)&(3,0)        & 0.0& 6.10 &no& 0 & \phantom{-}0.081 & -0.001 & 0.913 & \phantom{-}0.846  \\
              (4,0)&(3,0)      & 0.0& 6.10 & yes&  0  &  -0.649  &  -0.018  & 0.559&  \phantom{-}0.571  \\
                \hline
               (3,1)&(2,1)       &5.2& 3.81 &no& 2 & -0.060 & \phantom{-}0.019 & \phantom{-}0.619 & \phantom{-}0.934   \\
 
                 (3,1)&(2,1)        &5.2& 3.81& yes&  2 &\phantom{-}0.491   & -0.091    &  \phantom{-}0.409   &\phantom{-}0.604    \\
         
    \end{tabular}
    \caption{Comparison of the fitting parameters of the two heterostructures shown in Fig.~\ref{Fig:relax} for both the relaxed and the idealized (unrelaxed) structure. The first two lines describe the structure with a 3$\times3$ NbSe$_2$ supercell.}
    \label{Tab:relax}
\end{table}

\section{Details of Charge-to-spin calculations}
\label{App:UREEsupp}
\begin{figure*}
    \includegraphics[width=.99\linewidth]{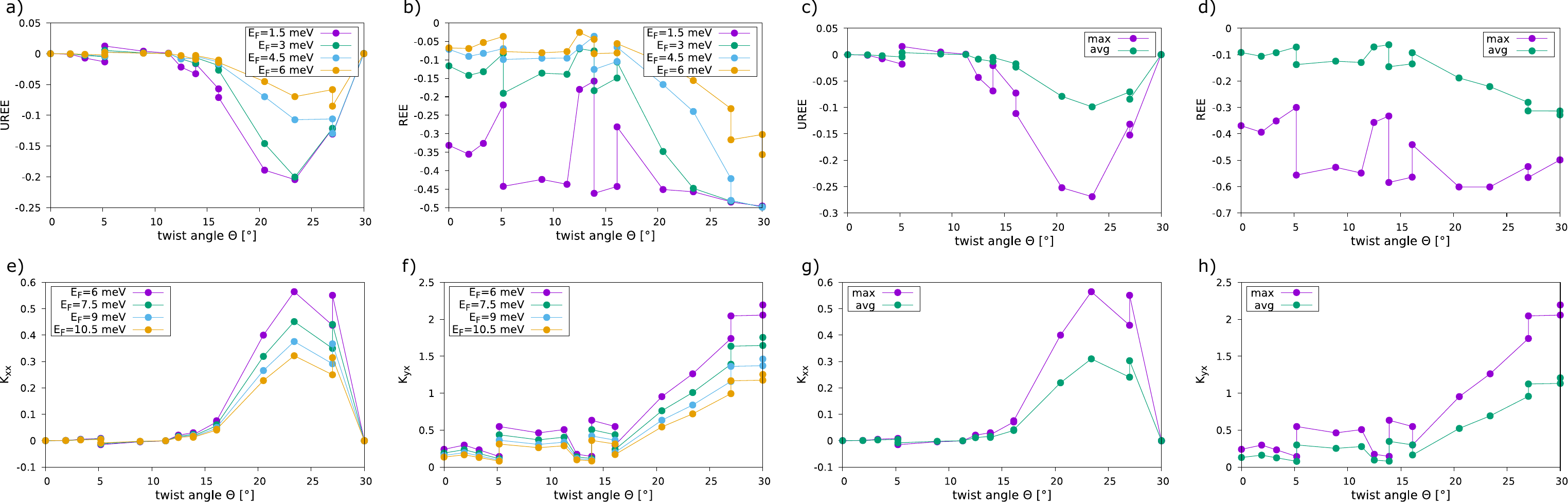}
    \caption{Comparison of the numerically calculated and analytical charge-to-spin convergence efficiencies as a function of twist angle.
    a) and b) show UREE and REE efficiencies respectively for a few selected Fermi levels. Each line can be seen as a cut through the data points of Fig.~\ref{Fig:UREE} d) or e) at a specific Fermi level. c) and d) also show the twist-angle dependency of the UREE and REE efficiencies respectively. Here, we plot two different quantities describing an overall efficiency independent of the Fermi level. The green line shows the efficiencies averaged over 12~meV ($E_F=0$ to $E_F=12~$meV, as in Fig.~\ref{Fig:UREE} c)), while the purple line shows the maximal efficiency that has been reached for any of the Fermi levels. e)-h) show the same as a)-d) only calculated by an alternative formula given in Ref.~\cite{Veneri22:PRB:radialrashba2} in arbitrary units. Instead of $\alpha_{UREE}$ and $\alpha_{REE}$, the spin responses are measured in $K_{xx}$ and $K_{yx}$ respectively. Here, the range of Fermi energies in which both the average and maximum are taken is 6~meV$<E_F<18$~meV instead of  0~meV$<E_F<12$~meV.\label{Fig:UREEsupp}}
\end{figure*}

In Sec.~\ref{Sec:UREE} we calculated efficiencies for the Rashba Edelstein effect (REE) and the unconventional Rashba Edelstein effect (UREE) using a Kubo formula approach. Since these efficiencies depend on both the twist angle and the Fermi energy, we plotted the data in a fenceplot (see Fig.~\ref{Fig:UREE} d) and e)). Here, we present alternative presentation forms. Fig.~\ref{Fig:UREEsupp} a) and b) show the data as cuts of different Fermi levels through the fence plot data. In Fig.~\ref{Fig:UREEsupp} c) and d) we plot two different measures for the total (U)REE (more precise: the value with maximal absolute value) of the (U)REE and the value averaged over 12~meV ($E_F=0$ to $E_F=12~$meV). The latter is the same as shown in Fig.~\ref{Fig:UREE} c).  For the UREE (Fig.~\ref{Fig:UREEsupp} c)) both measures convey the same message, since the total value is far from being capped by the maximally possible value. Taking into account only Rashba SOC, which is the source of both UREE and REE, this maximally possible value is $\alpha_{UREE}=\alpha_{REE}=0.5$. Other terms like $\lambda_{VZ}$ can bend the band structure in a way that flat bands give rise to a local maximum surpassing a certain Fermi level, while diminishing it for other Fermi levels. This can give a distorted picture of the truth, since in reality, fine tuning the Fermi level this precise might be impossible. Hence, for the REE (Fig.~\ref{Fig:UREEsupp} d)), the averaged curve gives a better overall picture of the true physics, as realizing an efficient REE device is easier with the broad high plateau of the 30$^\circ$ case. The drawback of plotting the averaged efficiency is that a range has to be specified over which the average has been taken. We assume that at $E_F=12$~meV the (U)REE has decayed enough and no new physics will emerge. Therefore this is the measure we use in the main paper.

We additionally use a different formula, which was given in Ref.~\cite{Veneri22:PRB:radialrashba2}, to calculate a measure for the charge-to-spin response of the system. It is derived within linear response theory as well and has an easy analytical form. It takes the same parameters, i.e. the parameters of the model Hamiltonian plus the Fermi level. The resulting quantities $K_{xx}$ and $K_{yx}$ can broadly be compared to the charge-to-spin conversion efficiencies ($\alpha_{REE}$ and $\alpha_{UREE}$) we calculate. However, they are not normalized by the charge density responses. The figure of merit $\varrho$ given in Ref.~\cite{Veneri22:PRB:radialrashba2} is more adequate to compare to and also has the correct sign. But since this would require information about the charge density response, we opt to use $K_{xx}$ and $K_{yx}$ instead. By showing the results in arbitrary units, we avoid the problem of having to give estimates for the other parameters of their model, which represent impurity strength and impurity density. Since we only want to compare the values of different twist angle, we deem this approach valid. We use the same set of parameters (determined by our fittings) as we used for the numerical approach. 
The results are shown in Fig.~\ref{Fig:UREEsupp} e)-h). Because the analytical formula is only valid for a Fermi level, where both subbands are already occupied, we change the range of Fermi levels to fulfil this condition for all twist angles (6~meV$<E_F<18$~meV).
For our set of parameter data we can conclude that both models qualitatively give the same intuitive results. However, there are two differences we can see:
\begin{enumerate}
    \item the analytical formula estimates the peak of the UREE to be slightly (about 3$^\circ$) to the right of the one determined by our numerical approach.
    \item the overall twist-angle tuneability is enhanced compared to the predictions of the numerical approach.
\end{enumerate}

\bibliography{references}

\end{document}